\documentclass[aip, preprint]{revtex4-1}

\usepackage{amsbsy,amssymb,amsmath,bm, gensymb}
\usepackage{graphicx,color,epsfig,rotate}

\begin{document}

\title{Probing the local environment of two-dimensional ordered vacancy structures  in Ga$_2$SeTe$_2$ via aberration-corrected electron microscopy}

\author{N.M. Abdul-Jabbar}
\affiliation{Department of Nuclear Engineering, University of California, Berkeley, California 94720, USA}
\affiliation{Materials Sciences Division, Lawrence Berkeley National Laboratory, Berkeley, California 94720, USA}

\author{P. Ercius}
\affiliation{National Center for Electron Microscopy, Lawrence Berkeley National Laboratory, Berkeley, California 94720, USA}

\author{R. Gronsky}
\affiliation{Department of Materials Science and Engineering, University of California, Berkeley, California 94720, USA}

\author{E.D. Bourret-Courchesne}
\affiliation{Materials Sciences Division, Lawrence Berkeley National Laboratory, Berkeley, California 94720, USA}

\author{B.D. Wirth}
\affiliation{Department of Nuclear Engineering, University of California, Berkeley, California 94720, USA}
\affiliation{Department of Nuclear Engineering, University of Tennessee, Knoxville, Tennessee 37996, USA}

\begin{abstract}

There has been considerable interest in chalcogenide alloys with high concentrations of native vacancies that lead to properties desirable for thermoelectric and phase-change materials. Recently, vacancy ordering has been identified as the mechanism for metal-insulator transitions observed in GeSb$_2$Te$_4$ and an unexpectedly low thermal conductivity in Ga$_2$Te$_3$. Here, we report the direct observation of vacancy ordering in Ga$_2$SeTe$_2$ utilizing aberration-corrected electron microscopy. Images reveal a cation-anion dumbbell inversion associated with the accommodation of vacancy ordering across the entire crystal. The result is a striking example of the interplay between native defects and local structure.

\end{abstract}

\maketitle

Ga$_2$SeTe$_2$ belongs to a class of III-VI semiconductors  with a defect zincblende structure (F$\bar{4}$3\textit{m} space group and lattice constant \textit{a} = 5.77 \AA) where, due to a valence mismatch, one third of the cation (group III element) sites are vacant. The high concentration of native vacancies in these materials is responsible for numerous interesting physical properties. Early investigations have shown that vacancies in these materials induce anomalously high radiation stability suitable for semiconductor devices operating in high-flux radiation environments \cite{Koshkin:vp, Koshkin:1976wj, Gurevich:1995wx}. Recently, there has been renewed interest in III-VI semiconductors for thermoelectric and phase-change random access memory applications. Ga$_2$Te$_3$ as a phase-change material, for example, shows better data retention ability, lower power consumption, and higher dynamic electric switching ratios when compared to the better-known Ge$_2$Sb$_2$Te$_5$ \cite{Zhu:2010jj}. Planar vacancy ordering in Ga$_2$Te$_3$ has been associated with strong phonon scattering, resulting in very low thermal conductivity, showing great potential as a thermoelectric material \cite{Kurosaki:2008cj}.

Theoretical and experimental investigations have shown that vacancy ordering in phase-change materials acts as a mechanism for metal-insulator transitions, whereby the transition to the metallic state is governed by the formation of ordered vacancy layers from the native vacancy clusters in the insulating state \cite{Siegrist:2011ej, Zhang:2012bp}. The result is an effective modification of electronic structure due to intrinsic structural rearrangements, rather than extrinsic compositional changes through doping. It follows that direct observation of such local structural changes, at the atomic level, is paramount for future discovery and understanding of a large class of materials with physical properties amenable to alteration by vacancy ordering. We attempt to do this at the required levels of spatial resolution by employing aberration-corrected scanning transmission electron microscopy (STEM) to image the atomic environment of Ga$_2$SeTe$_2$, with ordered two-dimensional vacancy structures.

Earlier electron microscopy of bulk Ga$_2$Te$_3$ and Ga$_2$Se$_3$ has revealed the presence of a mesoscopic superstructure of two-dimensional ordered vacancies within families of \{111\} planes \cite{Guymont:1992vj, Kienle:2003eg, Kim:2011kb}. In dark field images, these appear as dark streaks along [111] when viewed in $<$110$>$ zone axis orientations. Information about these defect structures via dark field imaging at the atomic level remains scarce. Consequently, we probe the environment in the vicinity of these defect structures (which also exist in Ga$_2$SeTe$_2$) by imaging in high-angle annular dark field (HAADF)-STEM utilizing aberration-corrected electron lenses, which drastically improve spatial resolution and image quality \cite{Spence:2003, Fultz:2007vo, Williams:2009uy}.

\begin{figure}[hb]
\includegraphics[width=12cm]{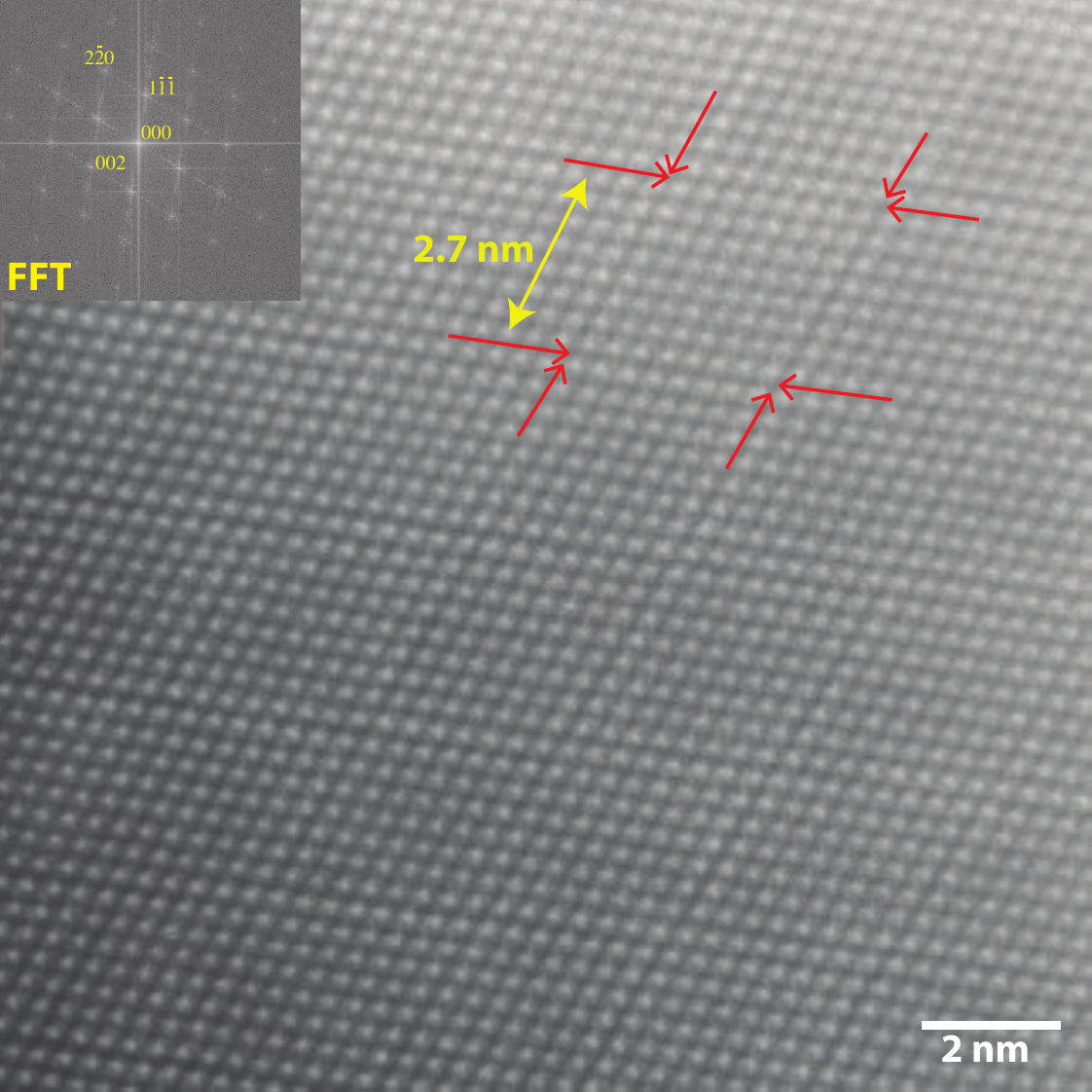}
\caption{HAADF-STEM image of bulk single crystal Ga$_2$SeTe$_2$ in the [110] zone axis, which reveals a periodic dark line contrast every eight \{111\} planes (arrowed). These planes contain a high concentration of vacancies arrayed in the highly ordered and self-assembled configuration displayed here.}
\label{fig1}
\end{figure}

\begin{figure}[ht]
\includegraphics[width=14cm]{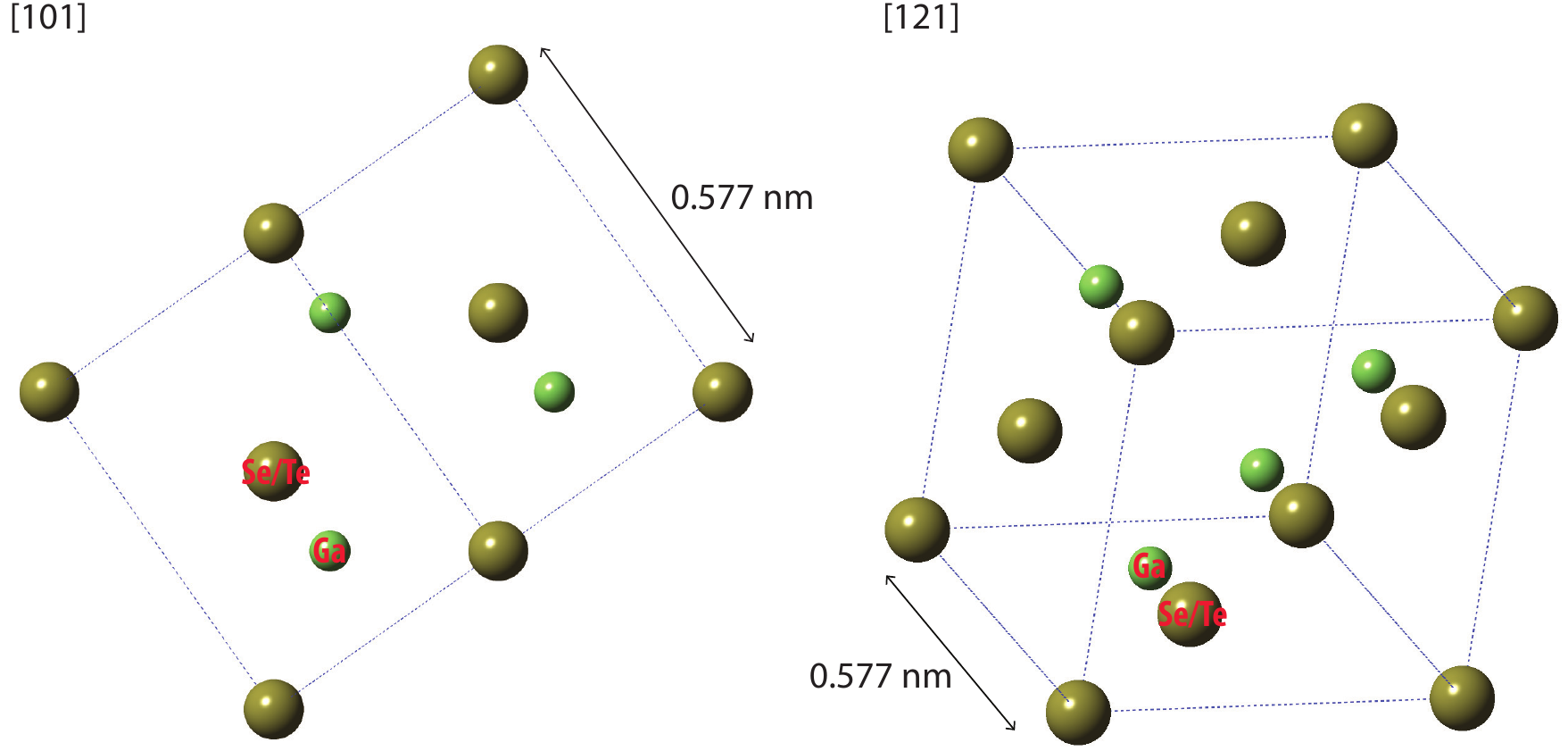}
\caption{Schematic of a Ga$_2$SeTe$_2$ crystal viewed in the [101] and [121] zone axes. }
\label{fig2}
\end{figure}

\begin{figure}
\includegraphics[width=12cm]{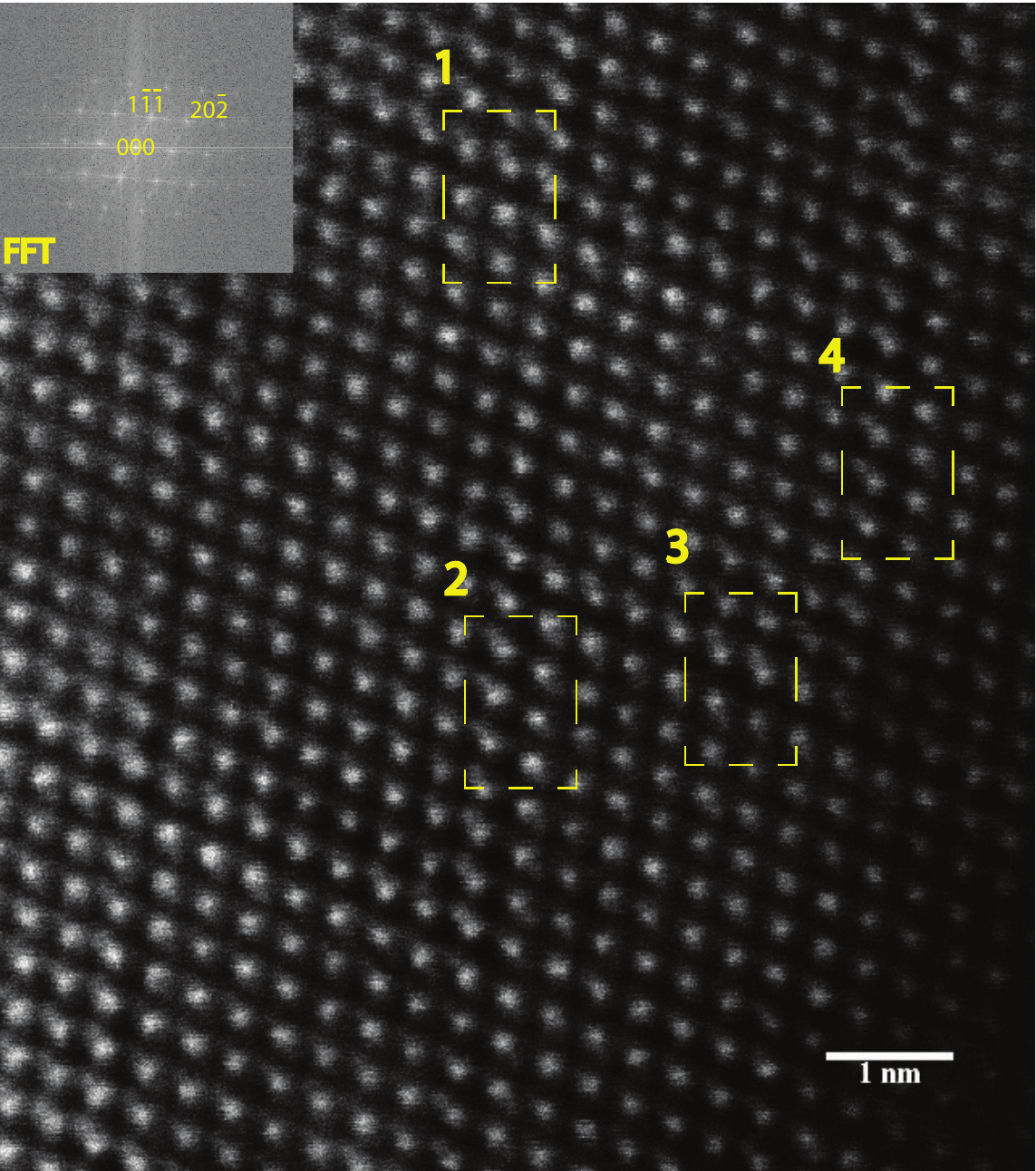}
\caption{HAADF-STEM image of bulk single crystal Ga$_2$SeTe$_2$ in the [101] zone axis at atomic resolution. As highlighted by the four regions, we observed cation-anion dumbbell inversions at both orientations of \{111\} planes.}
\label{fig3}
\end{figure}

\begin{figure}
\includegraphics[width=10cm]{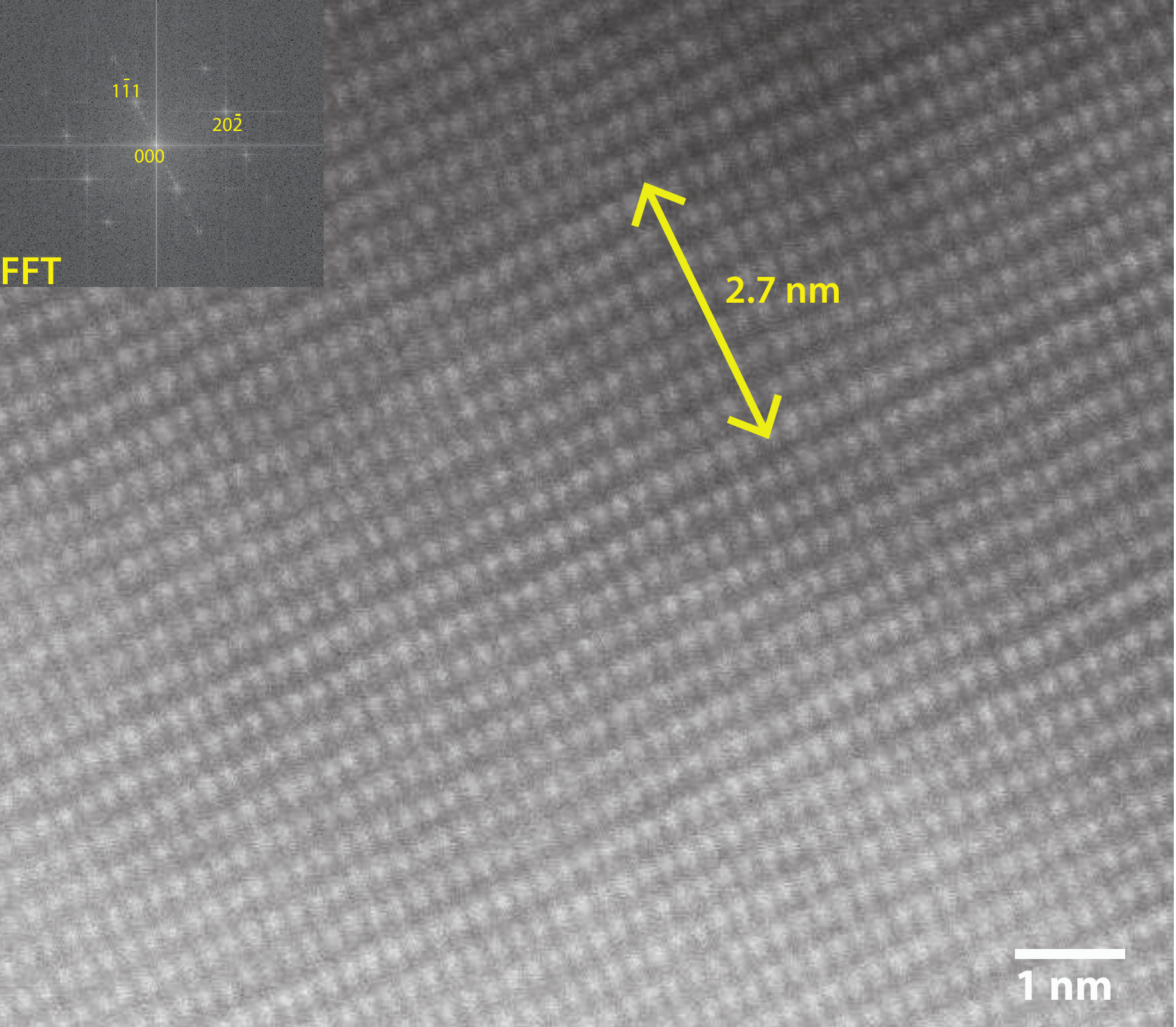}
\caption{HAADF-STEM image of bulk single crystal Ga$_2$SeTe$_2$ oriented in the [121] zone axis.}
\label{fig4}
\end{figure}

\begin{figure}
\includegraphics[width=8cm]{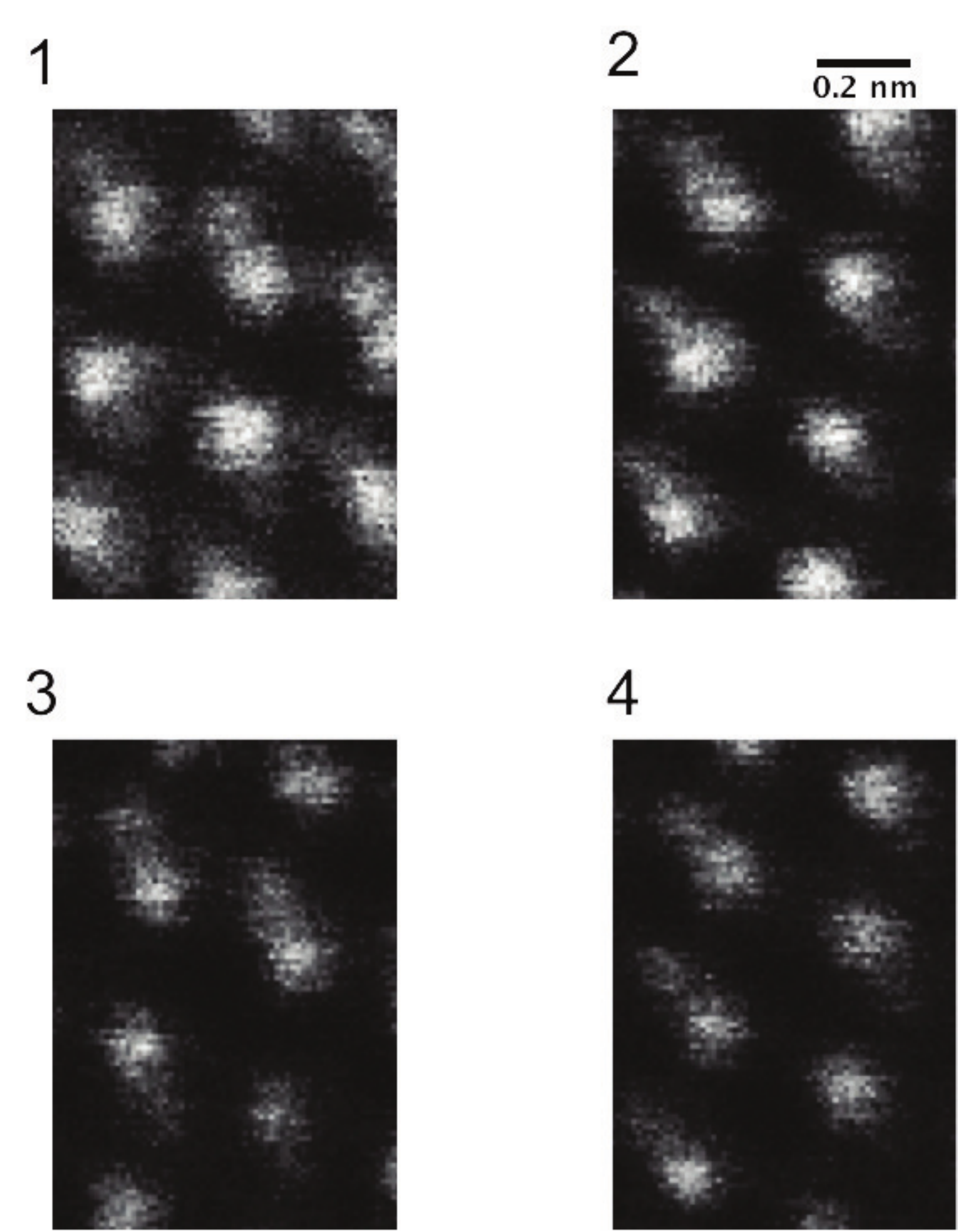}
\caption{Boundaries between vacancy-ordered planes in both $<$111$>$ directions highlighted in Figure 2 showing the cation-anion dumbbell inversion.}
\label{fig5}
\end{figure}

Figure 1 shows the atomic columns of a bulk Ga$_2$SeTe$_2$ crystal along a [110] projection. As expected from prior reports in the literature, we see a periodic contrast variation in the form of dark lines that are parallel to both sets of \{111\} planes visible in edge-on orientation along this zone axis. The inset Fourier transform of this image confirms the symmetry of atomic column stacking consistent with the face-centered cubic Bravais lattice associated with a zincblende crystal structure (illustrated in Figure 2) and the periodicity of the two-dimensional vacancy ordering.  The latter is evidenced by the fine spots at 1/8 the spacing of the \{111\} planes ($\approx$ 2.7 nm) along both $<$111$>$ reciprocal lattice directions in this projection.  

A magnified image of the Ga$_2$SeTe$_2$ crystal in the [101] zone axis is shown in Figure 3. Here, we are able to discern the cation-anion dumbbells that are expected in the zincblende lattice when viewed along the [101] projection; the dumbbells are also oriented perpendicular to the long axis created by the 20$\bar{2}$ reflection in reciprocal space (Figure 3 inset), which is the expected cation-anion orientation along $<$110$>$ projections. Based on the conventional mechanism of Z-contrast, we expect the brighter spots to represent the higher-Z anions (Te or Se) and the dimmer spots to represent the lower-Z cation (Ga) sites. The dumbbell distance is $\approx$ 0.147 nm in agreement with the predicted value. Additionally, we confirm the pervasiveness of \{111\} vacancy ordering in the zincblende structure by tilting the Ga$_2$SeTe$_2$ crystal to the [121] zone axis and collecting a dark field image (Figure 4). Consistent with their structural arrangement evident in $<$110$>$ zone axes, the modulation of vacancy-ordered \{111\} planes is again repeated at 2.7 nm intervals. However, spatially resolving both cation and anion sites ($\approx$ 0.085 nm) along $<$211$>$ zone axes is difficult due to deleterious effects arising from specimen thickness---where it is $\approx$ 15\% greater than $<$110$>$ zone axes thicknesses---and an amorphous layer that covers the crystal, which we had difficulty removing.

Examining the image detail in Figure 3 more carefully, we notice the dumbbell orientation in $<$110$>$ zone axes projections is inverted across vacancy-ordered planes in a manner analogous to an inversion twin boundary. Four yellow boxes in Figure 3 highlight such inversions. We note that the dumbbell inversion is constant within each domain bounded by the ordered vacancy-rich \{111\} planes. This illustrates a key feature of Ga$_2$SeTe$_2$ and its related compounds: native defect structures in these materials distort the local symmetry of the motif assigned to each lattice point, but the global symmetry of the Bravais lattice is preserved. This is captured in the STEM images, and confirmed by their inset Fourier transforms.

The mechanism driving the cation-anion dumbbell inversion across the vacancy-rich planes Ga$_2$SeTe$_2$ remains uncertain. The regular periodicity of this effect, however, suggests that the observed symmetry inversions are electronic in nature. Earlier x-ray investigations on Ga$_2$Te$_3$ have proposed that Ga vacancies may induce Jahn-Teller crystal distortions, where it was argued that unattached Te orbitals initiate a tetragonal distortion from the original cubic arrangement \cite{Otaki:2009cg, Otaki:2009if, Kashida:2009bc}. Our spatially-resolved structural microscopy cannot confirm the mechanism driving the observed local distortions. 

This investigation of Ga$_2$SeTe$_2$ serves as an example of a direct observation of the effect of native point defects on local structure. Our results show no detectable distortion in the lattice spanning the regularly periodic \{111\} planes along which vacancy ordering is concentrated;  however we do find an inversion in the cation-anion orientation vector across those planes, suggesting an electronic interaction at play. We conclude that further investigations of the atomic structure of Ga$_2$SeTe$_2$ and its related materials by modern characterization methods with high spatial resolution are key to developing further understanding of their technological potential.

\section{Methods}

Ga$_2$SeTe$_2$ single crystals were synthesized by mixing stoichiometric amounts of 8N Ga, 6N Se, and 6N Te in quartz crucibles sealed under vacuum (10$^{-6}$ Torr). The mixture was heated to 850 \degree C to form a sintered polycrystalline solid. The synthesized charge, still in the sealed crucible, was then placed in an 820 \degree C hot zone of a vertical Bridgman furnace, and translated through a temperature gradient of 10 \degree C /cm at 0.2 mm/h to a cold zone where solidification was complete. Ga$_2$SeTe$_2$ single crystals measuring 3 mm $\times$ 3 mm $\times$ 2 mm were harvested from the resulting ingot. A detailed procedure has been described elsewhere \cite{AbdulJabbar:2012bma}. To produce Ga$_2$SeTe$_2$ with ordered two dimensional vacancy structures, we annealed single crystal specimens in a sealed quartz crucible at 735 \degree C for 28 days, then quenched the sample to 0 \degree C. Single crystals were ground using an agate mortar and pestle mixed into an isopropanol solution. The solution was dispersed on gold ultra-thin carbon TEM grids for electron microscopy characterization. Electron microscopy was done at 300 kV using the TEAM I (a modified FEI Titan 80-300) microscope at the National Center for Electron Microscopy (NCEM) at Lawrence Berkeley National Laboratory (LBNL). 

\section{Acknowledgements}

We thank G. Gundiah (LBNL) and G-Y. Huang (UT Knoxville) for useful discussions. We also acknowledge M. Libbee (LBNL-NCEM) for her crucial role in sample preparation.  NMA acknowledges support from the Nuclear Nonproliferation International Safeguards Graduate Fellowship Program sponsored by the National Nuclear Security Administration's Next Generation Safeguards Initiative (NGSI). This work was performed at NCEM, which is supported by the Office of Science, Office of Basic Energy Sciences of the U.S. Department of Energy under Contract No. DE-AC02Ñ05CH11231. This work was supported by the US Department of Energy/NNSA/NA22 and carried out at the Lawrence Berkeley National Laboratory under Contract No.AC02Ñ05CH11231.


\end{document}